\documentclass{article}
\addtocounter{equation}{0}
\begin{document}
\title{Flux quantization for a vortex in two-gap superconductor}         
\author{E. \v{S}im\'{a}nek \footnote {Electronic address: simanek@ucr.edu}\\Department of Physics, University of California, Riverside, CA 92521}      
\date{}          
\maketitle

\begin{abstract}
Contrary to recent theoretical prediction, we show that the magnetic flux of a vortex in $SU(2)$ model of two-gap superconductor is quantized in units of $2 \pi/g$, not $4 \pi/g$. For the $U(1)$ version of this model, the flux is quantized in units of $2\pi\alpha/g$ where $0 < \alpha < 1$. The parameter $\alpha$ depends on the masses and concentrations of the Cooper pairs in the two condensates.
\end{abstract}

PACS number(s):  74.20.De, 74.25.Qt, 74.25.Sv, 74.90.+n
\\[10pt]

Ginzburg-Landau models with two flavors of Cooper pairs have attracted increased attention in recent years [1-3].  The discovery [4] of the two-band superconductor $MgB_{2}$ with surprisingly high transition temperature has sparked renewed interest in these models.  Starting from a Lagrangian for a charged doublet order parameter exhibiting $SU(2)$ symmetry, Cho [2] studied the flux quantization for a magnetic vortex.  In his approach , the magnetic flux is related to the boundary values of the gauge field at $r=0$ and $r= \infty$ where $r$ is the distance from the vortex center. The $r=0$ value is determined by examining the equation of motion for the magnitude of the order parameter.  Smothness of the order parameter at $r=0$ requires that this boundary value is nonzero.  Consequently, the flux is predicted in Ref. [2] to be quantized in the unit $4 \pi / g$ which is double of the flux carried by the well known Abrikosov vortex in one-gap superconductor [5].

In the present paper, we consider the same $SU(2)$ model of two-gap superconductor and calculate the magnetic flux for the vortex in terms of the Berry connection that is due to the texture induced by the vortex in  the charge space [2, 3]. In this way, we find that the vortex has a total flux quantized in the same unit as the Abelian vortex in one-gap superconductor.  Moreover, the present method allows us to consider the flux quantization in a more realistic $U(1)$ model of the two-gap superconductor.  In this model the flux appears to be quantized in the unit of $2 \pi/g$ times a parameter that can take any value in the interval $(0,1)$ dependent on the masses and concentrations of the Cooper pairs.

The Lagrangian density has the same form as that of Ref. [2]

\begin{equation}\label{Eq1}
\pounds= - |D_{\nu}\phi|^{2}+\mu^{2} \phi^{\dagger} \phi - \frac{\lambda}{2}(\phi^{\dagger} \phi)^{2}-\frac{1}{4}F^{2}_{\mu\nu}
\end{equation}

We consider a magnetic vortex in a (2 + 1)-dimensional spacetime.  For a time-independent configuration, $\phi$ is a function of two space variables (polar coordinates $(r, \varphi)$).  The object $D_{\nu} \phi$ is the covariant derivative

\begin{equation}\label{Eq2}
D_{\nu}\phi = (\partial_{\nu}+ igA_{\nu})\phi
\end{equation}

where $A_{\nu}$ is the Abelian gauge field given by

\begin{equation}\label{Eq3}
A_{\nu}= \frac{m}{g}A(r)\partial_{\nu}\varphi
\end{equation}

where $\varphi$ is the phase entering the vortex Ansatz $\phi(r, \varphi)$ parametrized by

\begin{equation}\label{Eq4}
\phi = \frac{1}{\sqrt{2}} \rho (r)\xi = \frac{1}{\sqrt{2}} \rho (r)\pmatrix{\cos \frac{f(r)}{2} \exp(-im\varphi)\cr \sin \frac{f(r)}{2}}
\end{equation}

where $f (r)$ decreases monotonically from $f (0) = \pi$ to $f (\infty) = 0.$

We note that expression (3) has a form dictated by requiring that the two terms in Eq. (2) cancel each other as $r\rightarrow \infty$ so that the first term in Eq. (1) yields a finite energy [6].

Varying $\pounds$ with respect to $A_{\nu}$ yields the gauge invariant current 

\begin{equation}\label{Eq5}
j_{\nu}= ig \Bigl[(D_{\nu}\phi)^{+} \phi -\phi^{\dagger}(D_{\nu}\phi)\Bigr]
\end{equation}

Using spinor (4), this equation takes the form

\begin{equation}\label{Eq6}
j_{\nu}= g^{2}\rho^{2}(r)(A_{\nu} + \tilde{A} _{\nu})
\end{equation}

where $\tilde{A}_{\nu}= - \frac{i}{g}\xi^{+}\partial_{\nu} \xi$ is the Berry connection characterizing the bundle of eigenstates (4). With the spinor $\xi$ defined in (4), we have

\begin{equation}\label{Eq7}
\tilde{A}_{\nu} = - \frac{m}{2g} \Big[1 + \cos f (r)\Big] \partial_{\nu}\varphi
\end{equation}

Let us consider Eq. (6) for a large circle $C$ of radius $R$ in the $xy$-plane.  For large $r$, the current density $j_{\nu}$ decays  exponentially owing to the Meissner effect.  Also the density $\rho (r)\rightarrow \rho_{0}$ as $r \rightarrow \infty$.  Thus, Eq. (6) implies that $A_{\nu} (R) \rightarrow - \tilde{A}_{\nu}(R)$ as $R \rightarrow \infty$.  Integrating this relation over $C$, we obtain with the use of the Stokes theorem the total magnetic flux passing through $C$

\begin{equation}\label{Eq8}
\hat {\Phi} = \int curl _{z} {\vec{A}} d^{2}x = \oint _{C} A_{\nu} d x _{\nu} \rightarrow - \oint _{C} \tilde{A}_{\nu} dx_{\nu}
\end{equation}

as $R \rightarrow \infty$. The right hand side (R. H. S.) of this equation can be evaluated using Eq. (7) and noting that $f (\infty) = 0$

\begin{equation}\label{Eq9}
\tilde{A}_{\nu} (R) = - \frac{m}{2g} \Big[1 + \cos f (R)\Big]  \partial _{\nu} \varphi \rightarrow  - \frac{m}{g} \partial_{\nu}\varphi
\end{equation}

as $R \rightarrow \infty$.  According to Eqs. (8) and (9), the flux is given by

\begin{equation}\label{Eq10}
\hat{\Phi} = \frac{m}{g} \oint _{C}\partial_{\nu} \phi d x _{\nu} = \frac{2 \pi m}{g}
\end{equation}

This result contrasts with the flux $\frac {4 \pi m}{g}$ obtained in Eq. (7) of Ref. [2].  Note that the gauge field $A(r)$ at the origin $r= 0$ does not appear in our derivation.

To understand this disagreement, we examine the curl of the gauge field (3) paying attention to the singularity at $r = 0$

\begin{equation}\label{Eq11}
curl_{z}\vec{A} = \varepsilon^{ \mu \nu} \partial _{\mu} A_{\nu} = \frac{m}{g}\frac{1}{r}\frac{dA(r)}{dr} + \frac{m}{g}A (r) \epsilon^{\mu \nu}\partial_{\mu}\partial _{\nu} \varphi
\end{equation}

The second term on the R. H. S. of this equation is proportional to $curl_{z} \frac {\vec{\varphi}_{0}}{r}$ where $\vec{\varphi}_{0}$ is the unit vector of the polar coordinates.  This quantity vanishes except along the line $r = 0$.  Applying the Stoke's theorem to a small circle around the origin, one obtains [7]

\begin{equation}\label{Eq12}
\epsilon^{\mu \nu} \partial _{\mu} \partial_{\nu} \varphi = curl_{z} \frac{\vec{\varphi}_{0}}{r} = 2 \pi \delta^{2} (\vec{r})
\end{equation}

Using Eqs. (11) and (12), the magnetic flux becomes

\begin{eqnarray}\label{Eq13}
\hat\Phi = \int curl_{z} \vec{A} d^{2}x = \frac{2 \pi m}{g} \Biggl[\int^{\infty}_{0} \frac{d A (r)}{dr} dr + \int ^{\infty} _{0} A (r) \delta ^{2} (\vec{r}) d^{2} x\Biggr] \nonumber \\* = \frac {2 \pi m}{g} \Bigg {\lbrace} \biggl[ A (\infty) - A (0) \biggr] + A (0) \Bigg {\rbrace} = \frac{2 \pi m}{g} A (\infty)
\end{eqnarray}

We see that the $A (0)$ term appearing in Eq. (7) of Ref. (2) is cancelled in Eq. (13) by the singular term stemming from Eq. (12). According to Eq. (5) of Ref. [2], we have $A(\infty) = 1$.  With this value, Eq. (13) predicts a flux $2 \pi m/g$ in agreement with Eq. (10).

In the  differential equation for $\rho (r)$ as derived in Ref. [2], the quantity $A (r)$ enters in the combination $[A (r) - \frac{\cos f + 1}{2}]^{2} \rightarrow A^{2}(0)$ as $r \rightarrow 0$.  This is to be compared with the form $[A (r)- 1]^{2}$ pertinent to the one-gap Ginzburg Landau superconductor.  Consequently, the requirement of smoothness of $\rho (r)$ at $r = 0$ imposes the boundary condition $A (r) \rightarrow - 1$ as $r \rightarrow 0$ [2]. 
     This contrast with the boundary condition $A (0)= 0$ for the one-gap case [6].

Ref. [2] proposes to remove this anomalous boundary condition by the gauge transformation

\begin{equation}\label{Eq14}
A_{\mu} \rightarrow A'_{\mu} = A_{\mu} + \frac{m}{g} \partial_{\mu} \varphi
\end{equation}

Recalling Eq. (3), this implies $A' (r) = [A (r) + 1]$ leading to new boundary values:  $A' (0)= 0$ and $A' (\infty) = 2$.  Using these values in Eq. (13), we obtain the gauge transformed flux $\hat{\Phi} = 4 \pi m/g$.  This does not imply, however, that the flux of the original vortex configuration $\Phi$ is quantized in units of $4 \pi/g$.  As pointed out in Ref. [7], the singular gauge transformation is not really a gauge transformation at the origin.  The term $\frac{m}{g} \partial _{\mu} \varphi$ on the R. H. S. of Eq. (14) injects a singular flux tube at $r=0$.  This is seen when one calculates the curl of this term as done in Eq. (12). The transformation of the gauge field given in Eq. (14) is accompanied by the transformation of the order parameter

\begin{equation}\label{Eq15}
\phi \rightarrow \phi' = \exp (-im \varphi) \phi
\end{equation}

Thus, whereas the vortex $\phi$ has a winding number $m$, the transformed vortex $\phi'$ has a winding number $2m$.  In both cases the flux remains to be quantized in units of $2 \pi/g$.

The flux quantization for the vortex Ansatz (4) can be also deduced by relating $\hat {\Phi}$ to the topological  (Pontryagin) index [6]

\begin{equation}\label{Eq16}
Q = \frac{1}{8 \pi} \int \epsilon^{\mu\nu} \vec{n}. (\partial _{\mu} \vec{n} \wedge \partial _{\nu} \vec{n}) d ^{2} x
\end{equation}

where  $\vec{n}= \xi^{+} \vec{\sigma} \xi$ is the unit vector field representing the local orientation in the charge space.  Applying Stoke's theorem to the R. H. S. of Eq. (8), we have

\begin{equation}\label{Eq17}
\hat{\Phi} = - \int \epsilon^{\mu\nu} \partial_{\mu} \tilde{A}_{\nu} d^{2}x
\end{equation}

With use of Eq. (4), we obtain $\vec{n} = (\sin f \cos m\varphi, \sin f \sin m\varphi, \cos f)$ yielding 

\begin{equation}\label{Eq.18}
\vec{n}. (\partial _{\mu}\vec{n}\wedge \partial_{\nu} \vec{n}) = m \sin f (\partial _{\mu} f  \partial_{\nu} \varphi - \partial _{\nu} f \partial_{\mu} \varphi)
\end{equation}

From the definition of the gauge field $\tilde{A}_{\mu}  = - \frac{i}{g}\xi^{+}\partial_{\mu} \xi $ we obtain using Eqs. (4) and (18)

\begin{equation}\label{Eq.19}
\epsilon^{\mu \nu} \partial_{\mu} \tilde{A}_{\nu} = \frac{1}{4g} \epsilon^{\mu \nu} \vec{n}. (\partial _{\mu} \vec{n} \wedge \partial _{\nu} \vec{n}) - \frac {\pi m}{g}(1 + {\cos}          f) \delta^{2}(\vec {r})
\end{equation}

where the second term arises from the singularity of $\vec{\partial} \varphi$ at $r = 0$ (see Eq. (12)).  Eq. (19) represents a generalization of the identity previously established for the case where zeros of the superfluid density (vortices) are absent [1],[8].  Using Eq. (19) in (17), we have 

\begin{equation}\label{Eq.20}
\hat{\Phi} = - \frac{1}{4g} \int \epsilon ^{\mu\nu} \vec{n}.  (\partial _{\mu} \vec{n} \wedge \partial _{\nu} \vec{n}) d^{2} x = - \frac{2 \pi}{g} Q
\end{equation}

Note that the second term of Eq. (19) does not contribute to Eq. (20) since the boundary condition $f (0) = \pi$ implies $1 + \cos f (0) = 0$.

The quantity $Q$ in Eq. (20) is the winding number which is the number of times the sphere $S^{2}_{field}$  is traversed  as we span the $R^{2}$ space compactified into the sphere $S ^{2}_{sp}$.  It should be noted that this compactification is enabled by the boundary condition $f (\infty) = 0$, implying 

\begin{equation}\label{Eq21}
\lim  _{r \to \infty}\vec{n}= (0,0, 1)
\end{equation}

In this way, one is led to an $R^{2}$ space with infinity identified which is homotopically equivalent to $S^{2}_{sp}$ [6].  This equivalence is established by stereographic projection from the north pole of $S^{2}_{sp}$ to $R^{2}$.  The mappings $S^{2}_{sp} \rightarrow S^{2}_{field}$ are representations of the second homotopy group $\pi_{2} (S^{2}) = Z$ where $Z$ is the group of integers under addition [6].  Hence the degree of these mappings is $Q = 0 , \pm 1, \pm 2$ ... The specific value  of $Q$, consistent with the vortex Ansatz (4), is obtained by evaluating the local topological density in Eq. (16) from Eq. (18) in cylindrical coordinates

\begin{eqnarray}\label{Eq22}
\frac{1}{8 \pi} \epsilon ^{\mu \nu} \vec{n}. (\partial _{\mu} \vec{n} \wedge \partial _{\nu} \vec{n}) = \frac {m}{8 \pi} \sin f \epsilon ^{\mu \nu} (\partial _{\mu} f \partial _{\nu} \varphi - \partial _{\nu} f \partial _{\mu} \varphi) \nonumber \\* = \frac {m}{4 \pi r} \sin f \frac {df}{dr} = - \frac {m} {4 \pi r}\frac { d \cos f}{dr}
\end{eqnarray}

Using this result in Eq. (16), we obtain

\begin{equation}\label{Eq23}
Q = - \frac {m}{2} \int ^{\infty}_{0} \frac {d \cos f}{d r} d r = - \frac {m}{2} [\cos f (\infty) - \cos f (0)] = - m
\end{equation}

where we used the boundary conditions $f (0) = \pi$ and $f (\infty) = 0$. Eqs. (20) and (23) imply $\hat \Phi = \frac {2 \pi m}{g}$ in agreement with Eq. (10).

In conclusion, the present calculations show that the magnetic flux of the vortex in the $SU (2)$ symmetric model of two-gap superconductor is quantized in the unit of $\frac {2 \pi}{g}$.  For condensates formed by Cooper pairs, the coupling constant $g$ can be related to the single electron charge $e$.  We have $g = \frac {2e}{\hbar c}$ so that the unit $\frac { 2 \pi}{g} = \frac {h c}{2e} = \Phi_{0}$ where $\Phi_{0}$ is flux quantum carried by Abrikosov flux tubes [5].

The $SU(2)$ rotational symmetry of the Lagrangian may not be easily achieved in real samples since it requires that the coupling constants of the potential term satisfy rather restrictive conditions [2, 3].  A more realistic potential energy has been considered in Ref. [1].  It has the generic form 

\begin{equation}\label{Eq24}
V = A + B n_{3} + C n^{2} _{3}
\end{equation}

where $A, B$ and $C$ are functions of masses and Copper pair (24) densities.  The minimum of $V$ takes place for $n_{3} = \tilde{n}_{3} = - \frac{B}{2C}$.  Hence, the ground state value of the vector $\vec{n}$ is a circle on the $S^{2}_{field}$ sphere.  Now the $SU(2)$ symmetry of the Lagrangian is broken to $U (1) \times U (1)$ symmetry.  The two $U(1)$ symmetries correspond to the conservation of integrals of the two quantities, $N_{1} + N_{2}$ and $N_{1} - N_{2}$, where $N_{i} = |\Psi_{i}|^{2}$ is the density of the $i-$th condensate. 

If the potential in Eq. (1) is replaced by the $U(1)$ symmetric form (24), the boundary condition (21) is replaced by

\begin{equation}\label{Eq25}
\lim _{r \to \infty} \vec{n} = \Bigl{(}\sqrt{1-\tilde{n}^{2}_{3}} \cos \varphi, \sqrt {1 - \tilde{n}^{2}_{3}} \sin \varphi, \tilde {n}_{3}\Bigr{)} 
\end{equation}

This implies that the infinity of the $R^{2}$ space cannot be identified so that this space cannot be compactified into a sphere $S^{2}_{sp}$.  Instead of the mapping $S^{2}_{sp} \rightarrow S ^{2}_{field}$, we have a map of $R^{2}$ to a fraction  of the sphere  $S^{2}_{field}$ corresponding  to the interval $(n_{3} = - 1, \tilde{n}_{3})$.  Consequently, the topological index $Q$ becomes a fraction of $m$.  Using $\cos f (\infty) = \tilde {n}_{3}$ and $\cos f (0) = - 1$, Eq. (23) yields

\begin{equation}\label{Eq26}
Q = - \frac {m}{2} (\tilde {n}_{3} + 1)
\end{equation}

so that the flux (20) becomes 

\begin{equation}\label{Eq27}
\hat {\Phi} = \frac {2\pi}{g} m  \Bigl{(}\frac {1 + \tilde {n}_{3}}{2}\Bigr{)}
\end{equation}

Consequently, the vortex in the $U(1)$ model of the two-gap superconductor is quantized in units $\frac{2 \pi}{g} \alpha$, where $\alpha = \frac{1 +\tilde{n}_{3}}{2}$ is a continuous parameter satisfying the inequality $0 < \alpha < 1$.

According to Ref. [1], the parameter $B$ vanishes when the condensates satisfy the conditions $m_{1} = m _{2}$ and $N_{1} = N_{2}$.  In this case, the equilibrium value of $n_{3}$ is $\tilde {n}_{3} = 0$ and Eq. (26) leads to a topological index $Q = - m/2$.  Hence, we have a soliton which has half the winding number of the $SU(2)$ vortex (meron).  At the infinity, the vector $\vec{n}$ in the meron configuration points out radially in the $xy$-plane like the magnetization of a magnetic vortex in an easy plane ferromagnet [9]. We note that a soliton with $Q = m/2$ is also possible.  Eq. (23) shows that this takes place when $f (\infty) = \pi/2$ and $f (0) = 0$.  However, the boundary condition $f (0) = 0$ requires a vortex Ansatz that is reasonable at the north pole of the $S^{2}_{field}$ sphere [10].  It is obtained by multiplying Eq. (4) by $\exp (im \varphi )$ thus reversing the winding number.  Soliton with topological index $Q = \pm m/2$ is reminiscent of the meron excitations in the $U(1)$ symmetric case in double-layer systems studied previously by Moon et al. [11]. In these systems the z-component of the pseudospin density at asymptotically large distances from the vortex center vanishes in order to minimize the charging energy of the capacitor formed by the two layers.  This is formally analogous to our case where $\tilde {n} = 0$ follows by minimizing the potential energy (24) where the parameter $B$ is set equal to zero.  However, the condition of symmetric bands $(m_{1} = m_{2}, N _{1} = N_{2})$ does not seem satisfied in real materials.  For instance, Ref. [12] confirms the existence of two different pair potentials and spatial scales for the two bands in $MgB_{2}$ superconductor.


\begin{thebibliography}{5}

\bibitem{1}
E. Babaev, L.~D. Faddeev, and A.~J. Niemi,
\emph{}
Phys. Rev. B \textbf{65}, 100512 (2002).


\bibitem{2}
Y.~M. Cho,
\emph{}
Phys. Rev. B \textbf{72}, 212516 (2005).


\bibitem{3}
Y.~M. Cho, H. Khim, and N. Yong,
\emph{}
cond-mat/0308182 (unpublished). 

\bibitem{4}
F. Bouquet et al, 
\emph{}
Phys. Rev. Lett. \textbf{87}, 047001 (2001); Amy Y. Liuet et al, ibid \textbf{87}, 087005 (2001); P. Szabo et al, ibid \textbf {87}, 137005 (2001). 

\bibitem{5}
A. Abrikosov, 
\emph{}
Sov. Phys. JETP \textbf{5}, (1957).

\bibitem{6}
V. Rubakov,
\emph{Classical Theory of Gauge Fields,}
(Princeton University Press, Princeton and Oxford, 2002).


\bibitem{7}
K. Bardakci and S. Samuel,
\emph{}
Phys. Rev. D \textbf{18}, 2849 (1978).


\bibitem{8}
E. Babaev, 
\emph{}
Phys. Rev. Letters \textbf{88}, 177002 (2002).

\bibitem{9}
M.~C. Ogilvie, and G.~S. Guralnik, 
\emph{}
Nuclear Physics B \textbf{190}, 325 (1981).


\bibitem{10}
M. Stone, 
\emph{}
Phys. Rev. D \textbf{33}, 1191 (1986).


\bibitem{11}
K. Moon et al, 
\emph{}
Phys. Rev. B \textbf{51}, 5138 (1995).

\bibitem{12}
A.~E. Koshelev and A. A. Golubov,  
\emph{}
Phys. Rev. Letters \textbf{90}, 177002 (2003).





\end{thebibliography}
\end{document}